\documentclass[12pt]{article}
\usepackage[T1]{fontenc}
\usepackage[francais,english]{babel}
\usepackage[dvips]{graphicx}
\usepackage{amsfonts}
\usepackage{latexsym}
\usepackage{mathptmx}
\usepackage{amsmath}
\usepackage{lscape}
\usepackage{longtable}
\usepackage{lscape}
\usepackage{rotating}
\usepackage{longtable}
\usepackage{setspace}
\usepackage{anysize}

\marginsize{.9in}{.9in}{1.5in}{1.5in}

\def\picture #1 by #2 (#3){\dimen0 = \hsize
\advance\dimen0 by -#1 \divide\dimen0 by 2 \hskip
\dimen0 \vbox to #2{\hrule width #1 height 0pt depth 0pt
\vfill \special{picture #3}}}

\def \R{{\mathbb R}}

\font\ninerm=cmr9
\long\outer\def\abstract#1{\bigskip\vbox{\noindent\ninerm
\baselineskip=10pt#1}\nobreak\bigskip}
\catcode`\â= \active \def â{\^a}
\catcode`\ê= \active \def ê{\^e}
\catcode`\ô= \active \def ô{\^o}
\catcode`\û= \active \def û{\^u}
\catcode`\î= \active \def î{{\^\i}}
\catcode`\é= \active \def é{\'e}
\catcode`\è= \active \def è{\`e}
\catcode`\à= \active \def à{\`a}
\catcode`\ù= \active \def ù{\`u}
\catcode`\ç= \active \def ç{\c {c}}
\catcode`\ï= \active \def ï{{\"\i}}

\newcount\numero
\def\exo#1{\advance\numero by 1\bigskip
{\noindent\tenbf #1\the\numero. }}
\def\frac#1#2{{#1\over #2}}
\numberwithin{equation}{section}

\begin{document}

\title{Viewing Risk Measures as Information}

\author{Dominique Guégan\footnote{Paris School of Economics, MSE - CES, Université Paris1 Panthéon-Sorbonne, 106 boulevard de l'hopital, 75013 Paris, France, e-mail: dominique.guegan@univ-paris1.fr}
, Wayne Tarrant\footnote{Department of Mathematics, Wingate University, Wingate, NC 28174, USA, e-mail: w.tarrant@wingate.edu}}

\maketitle

\begin{abstract}

\noindent \textit{Abstract}:  Regulation and risk management in banks depend on underlying risk measures. In general this is the only purpose that is seen for risk measures. In this paper we suggest that the reporting of risk measures can be used to determine the loss distribution function for a financial entity. We demonstrate that a lack of sufficient information can lead to ambiguous risk situations. We give examples, showing the need for the reporting of multiple risk measures in order to determine a bank's loss distribution. We conclude by suggesting a regulatory requirement of multiple risk measures being reported by banks, giving specific recommendations.\\



\end{abstract}

\vfil \eject

\section{Introduction}

\noindent In every instance that we can find, risk measures are used to compute an amount of capital that an institution should hold in order to remain financially solvent, whether this calculation is performed for a bank's own internal risk department or for the requirements of an external entity. For instance regulatory bodies like the Basel Committee on Banking Supervision require the reporting of the Value at Risk (VaR) in order to legislate the amount of cash reserves a bank must have. But should this be the only function of a risk measure? \\

\noindent In this paper we present the thesis that the reporting of a risk measure is the revelation of a piece of information about the financial health of an institution. We show that there are multiple possibilities for loss distributions when only a VaR is reported. We then introduce two other measures, the Expected Shortfall and the Maximum Loss. We show that reporting any one of these measures can lead to multiple possible loss distributions. \\

\noindent Following the old adage that more information is always better, we consider the situation of reporting any two risk measures. Again, we show that there is ambiguity about the loss distribution, no matter which two measures we choose. We show the same result for three measures and for five measures. We then conclude with some suggestions for regulators and with some suggestions for further research.

\section{How can we measure risk?}

\noindent There is little wonder that people are confused by what the term ``risk'' ought to mean. For instance, the website businessdictionary.com defines risk in seventeen different ``general categories''. Now, each of the seventeen definitions are relevant to the situations of banks and securities firms. The unifying theme for each of the definitions is that risk requires both uncertainty and exposure. If a company already knows that a loan will default, there is no uncertainty and thus no risk. And if the bank decides not to loan to a business that is considered likely to default, there is also no risk for that bank as the bank has no exposure to the possibility of loss. \\

\noindent If we desire to measure risk in some way, one goal could be to have a single risk measure that will account for all the risk that a bank or securities firm might encounter. Some have objected to the risk measure being a single number, but there is some support for this idea. Investing is always a binary decision- either one invests or one chooses not to invest. Thus the argument is that, given a single number, one should have enough information to decide whether to invest or not. There have been some general agreements about the kinds of properties that such a risk measure ought to possess. \\

\noindent Virtually everyone would agree that if the payout for an investment is always positive (after accounting for the risk-free interest rate), then there is truly no risk of loss in the investment. Many people would say that there is ten times as much risk in investing \$10,000 as there is in investing \$1000 simply because there is ten times as much money at stake. And most would acknowledge that having cash on hand makes them feel safer about making an investment. It was exactly these thoughts that were, respectively, codified into the definition of risk measure. \\

\noindent Let $X$ be a random variable. Then $\rho$ is a risk measure if it satisfies the following properties:
\begin{itemize}
\item (Monotonicity) if $X \ge 0$, then $\rho(X) \le 0$
\item (Positive Homogeneity) $\rho(\alpha X) = \alpha \rho(X) \ \forall \ \alpha \ge 0$
\item (Translation Invariance) $\rho(X + a) = \rho(X) - a \ \forall \ a \in \R$.
\end{itemize}

\noindent The idea behind this definition is that a positive number implies that one is at risk for losing capital and should have that positive number of a cash balance on hand to offset this potential loss. A negative number would say that the company has enough capital to take on more risk or to return some of its cash to other operations or to its shareholders. \\

\noindent The risk measure that is most used is the Value at Risk (VaR). Essentially the $\alpha$-Value at Risk is that number L so that we can expect the losses to be worse than L exactly 1 - $\alpha$ of the time. For instance if the 95\%- Value at Risk of our position is \$100, we would expect to lose more than \$100 only 5\% of the time. A more formal definition can be stated as follows: \\

\noindent The $\alpha$- Value at Risk of a position $X$, $VaR_{\alpha}(X) = -inf \lbrace x: P(X>x) \le 1 - \alpha \rbrace$ \\

\noindent Since the 1951 paper of Markowitz, many have also favored the usefulness of diversification. VaR does not account for this preference, and a simple example shows this. Consider the case where a bank has made two \$1 million loans and one \$2 million loan, each with a 0.04 probability of default and all pairwise independent. Then the 95\%-VaR for each loan is \$0. Thus, if we construct a portfolio consisting solely of the \$2 million loan, we must have a 95\%-VaR of \$0. If we instead choose diversification and make our portfolio out of the two independent \$1 million loans, something paradoxical happens. The probability of both loans defaulting is 0.0016, but the probability of exactly one loan defaulting is 0.0768. This implies that the 95\%-VaR of our diversified portfolio is \$1 million. Thus, VaR does not favor diversification.\\

\noindent This leads Artzner, Delbaen, Eber, and Heath (1997) to term VaR as incoherent. They define a coherent risk measure as a risk measure that also favors diversification in the following way:

$$
{\rm if} \; \; X_1, X_2\;  {\rm are}\;  {\rm random}\; {\rm variables,}\; {\rm then} \; \;  \rho(X_1 + X_2) \le \rho(X_1) + \rho(X_2)
$$

\noindent The most commonly used coherent risk measure is the Expected Shortfall. The $\alpha$-Expected Shortfall is the amount one would expect to lose on average, given that one is in a case where the loss is greater than the $\alpha$-VaR. Specifically, the $\alpha$-Expected Shortfall of a position $X$, $ES_{\alpha}(X)$, is $ES_{\alpha}(X) = -E[x] | x < -VaR_{\alpha}(X)$. \\

\noindent Finally, most risk managers want to account for the worst case scenario. Indeed it is only prudent to understand what could be at stake before agreeing to take a position. This leads to the introduction of one final risk measure, called the Maximum Loss. Given a position X, the Maximum Loss of a position $X$, $ML(X)$, is the worst loss one could experience with nonzero probability, ie, $ML(X) = -inf \lbrace (X) | p(X) > 0 \rbrace$. \\

\section{Ambiguity left by the reporting of risk measures}

We note that in the rest of this paper we consider a single timeframe for all the risk measures under consideration. That is to say that all calculations are assumed to be over the identical historical timeframe. For example we might consider a 100-day 95\% VaR and a 100-day 95\% ES, but in this section we will not consider a 100-day 95\% VaR and a 50-day 95\% ES. Of course the Maximum Loss could be an historical one, though this is often computed from a model of expected returns or losses.

\subsection{Insufficiency of a single measure}

\noindent Here, we discuss the weakness of any single measure of risk for risk management strategy. \\

\noindent In the following we are concerned with tails of probability distribution functions, ie, if $f(x)$ is the pdf of a loss distribution, then we are concerned with the region $\lbrack c,d \rbrack$ with the property that $\int_c^d f(x) dx = 0.05$ and that $f(x) = 0$ for all $x>d$. In Figure 1, $\lbrack c,d \rbrack$ is the region of the graph that we have shown. This is often called the tail of the loss distribution. In Figure 2 and following, we will only see the region $\lbrack c,d \rbrack$. Under our notation, the 95\%-VaR($X$) = $c$, the $ML=d$, and the 95\%-ES($X$) = $\frac{1}{.05} \int_c^d x f(x) dx $. We also want to mention that the probability density function of the loss is just the negative of the probability density function for the returns.\\

\begin{figure}[h!]
\centering%
\includegraphics{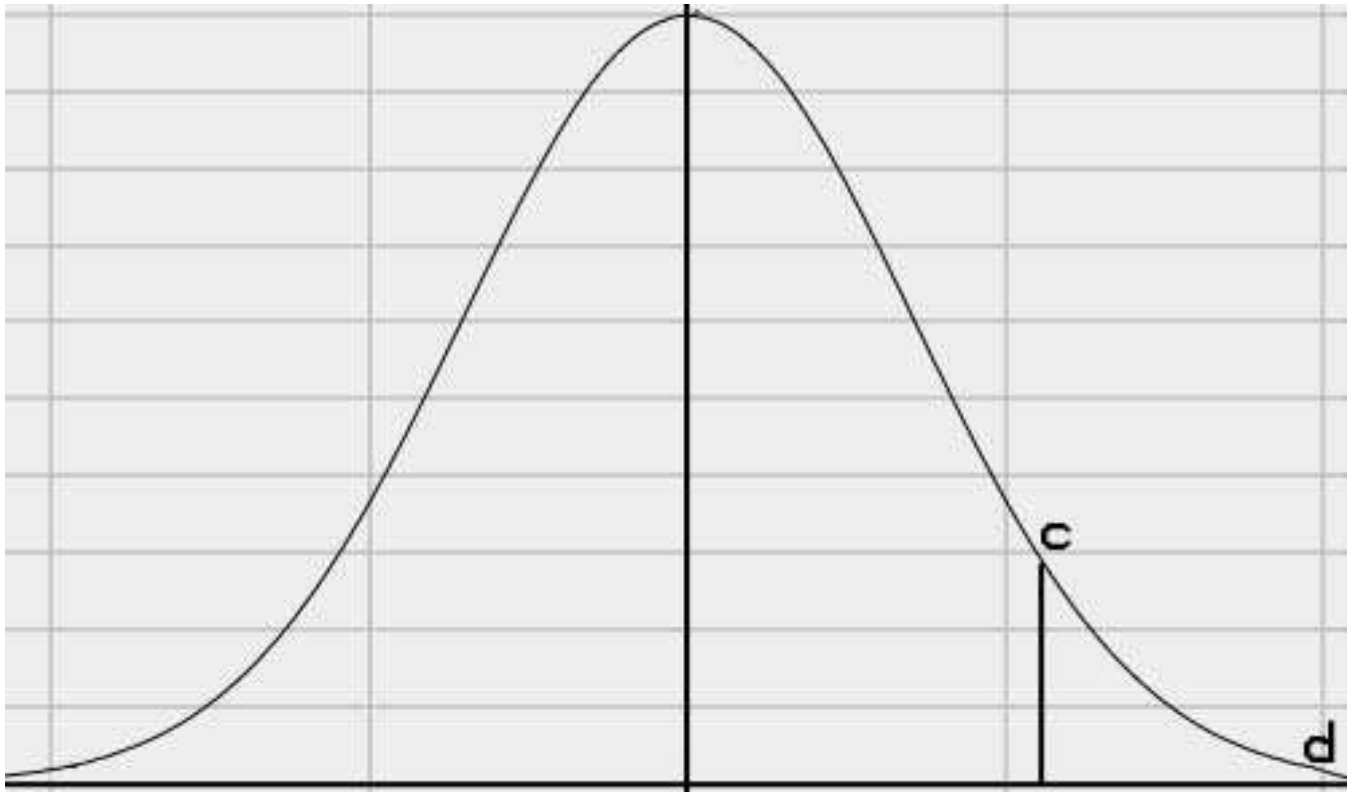}
\caption{The region of interest [c,d] in a loss pdf}
\end{figure}

\noindent In this and the following sections we will also adopt the notation that the first graph in each figure represents a loss distribution $X_1$, the second represents a loss distribution $X_2$, the third a loss distribution $X_3$, and so on. There is no significance to this. It is purely for ease of exposition.\\

\noindent Although Value at Risk is used by the Basel Committee for determining the capitalization that a bank needs, it turns out that VaR is insufficient for distinguishing between some kinds of different risk characteristics. Assume that a bank reports that it has a 95\%-VaR of 0 for its present balance sheet. \\

\begin{figure}[h!]
\centering%
\includegraphics{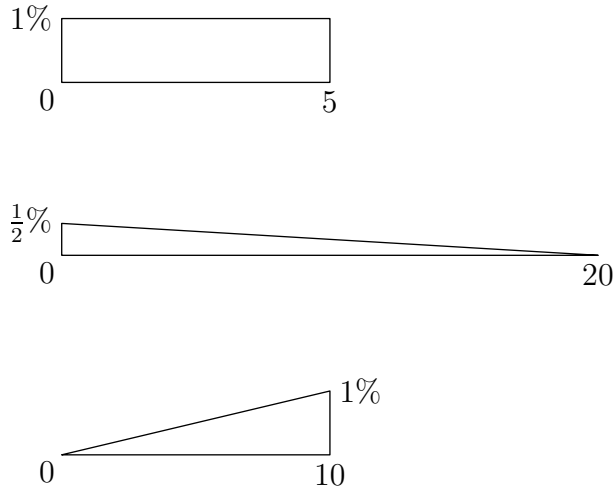}
\caption{Loss distributions with 95\%-VaRs equal to 0}
\end{figure}

\noindent The bank is able to have any of the tails in Figure 2 as the tail of its loss distribution. In each case, the 95\%-Value at Risk is $0$. This is a great misnomer, as there are probabilities of a loss in each case. We see that the loss distribution $X_1$ gives a uniform probability of loss from the VaR of 0 out to 5. The loss distribution $X_2$ is a very standard tail of a probability distribution, with the probability of losses decreasing as the size of losses increase. The position $X_3$ is something that might look strange at first glance. However, such loss distributions may occur when risks are highly concentrated. For instance we have seen increasing probability of larger and larger losses in the real estate market in some American cities. As one house is foreclosed upon and sits empty, it is a blight to a neighborhood, causing other houses to fall in value and become more likely to enter foreclosure. Since community banks often focus their loans in one particular locality, there is a potential for such a distribution to occur. \\

\noindent It is notable that our other two measures can help us distinguish among the three loss distributions. We can calculate that $ES_{95\%}(X_1) = \frac{5}{2}$, $ES_{95\%}(X_2)= \frac{20}{3} = ES_{95\%}(X_3)$, so we are unable to tell the difference between the second and third loss distributions. However, we have that $ML(X_1) = 5$, $ML(X_2) = 20$, and $ML(X_3) = 10$, which distinguishes each of these three loss distributions from the other two. \\

\noindent Whether one argues that certain loss distributions are likely or unlikely, there is a stubborn fact that remains. The Value at Risk cannot distinguish among uniform, rising, and falling probabilities. Since most bank regulations only require the reporting of VaR, this should be a very troubling fact.\\

\noindent Since VaR turned out to be insufficient for our risk measuring purposes, maybe one of the two other measures is a better choice. We will start with the Expected Shortfall at the 95\% level. \\

\begin{figure}[h!]
\centering%
\includegraphics{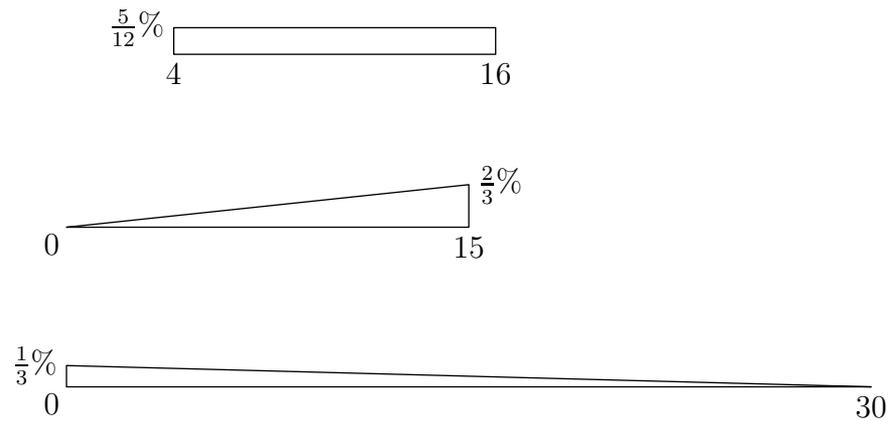}
\caption{Loss distributions with 95\%-ESs of 10}
\end{figure}

\noindent In Figure 3 we see that the reporting of the 95\%-Expected Shortfall leaves the same issue as before. There are uniform, rising, and falling probabilities, all of which have the same 95\%-Expected Shortfall of 10. In this case, the Maximum Loss once again identifies the three situations because $ML(X_1) = 16$, $ML(X_2) = 15$, and $ML(X_3) = 30$. Here we also have that $VaR(X_1) = 4$, while $VaR(X_2) = 0 = VaR(X_3)$, where each of the VaR calculations are at the 95\% level.\\

\noindent Because the ES has not been able to distinguish among very different risk situations, we turn to the other measure we have introduced - the Maximum Loss. We will again look at the last 5\% of the probability distribution so that we might be able to consider the usefulness of our other two measures.\\

\begin{figure}[h!]
\centering%
\includegraphics{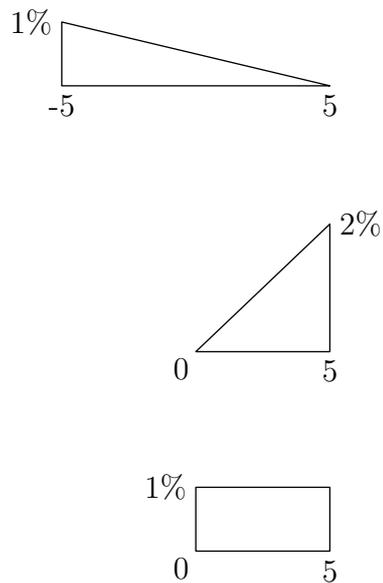}
\caption{Loss distributions with MLs of 5}
\end{figure}

\noindent As we see in Figure 4, the risk models have the same Maximum Loss as all the distributions have no probability occurring after the value of $5$. Again, there are very different situations of risk here though. The first loss distribution is such that $VaR(X_1) = -5$. For the second and third loss distributions we have that $VaR(X_2) = 0 = VaR(X_3)$. And this time the Expected Shortfall sorts out the different distributions because $ES(X_1) = -\frac{5}{3}$, $ES(X_2) = \frac{10}{3}$, and $ES(X_3) = \frac{5}{2}$.\\

\noindent We hope that the geometry of our figures has pointed out some facts about the different risk measures. In order for us to obtain two distributions with the same 95\%-Value at Risk, we only have to begin our loss distribution tails at the same value, making sure that we have 5\% of the probability after that common point. For two distributions to have the same 95\%-Expected Shortfall, we must have two geometric figures that have the same weighted average. And for two figures to have the same Maximum Loss, we would require a common point as the maximum value for which both loss distributions have a nonzero probability.

\subsection{Insufficiency of any two risk measures}

\noindent Since individual measures failed to give a thorough picture of the risk characteristics of a given loss model, maybe the situation will be better if we try using two different risk measures. We will attempt to use measures in pairs to see if they give us a full description of the risk characteristics of a loss model.\\

\noindent We will start by using both the 95\%-VaR and the 95\%-ES. Again we are left with a problem. In Figure 5 we see that the loss distribution $X_1$ is uniform, while $X_2$ has rising probabilities as losses grow, and $X_3$ is the most standard looking probability tail. It is discouraging that the two most frequently used measures do not distinguish among these three situations. Each of the distributions has a 95\%-VaR of 0 and each of them has an 95\%-ES of 10. With such different loss characteristics sharing the same VaR and ES, it is astounding that these are the only two measures mentioned in the Basel 3 document. We are encouraged, though, because here the Maximum Loss will distinguish among the loss distributions as $ML(X_1) = 20$, $ML(X_2) = 15$, and $ML(X_3) = 30$.\\

\begin{figure}[h!]
\centering%
\includegraphics{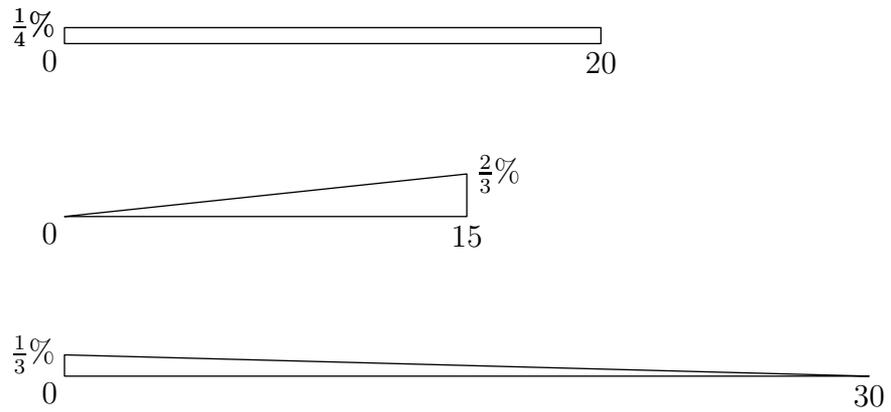}
\caption{Loss distributions with equal 95\%-VaRs and equal 95\%-ESs}
\end{figure}

\noindent Now we try to use the pair of measures that are the 95\%-VaR and the Maximum Loss. As shown in Figure 6, we can still have some ambiguity, as in previous examples. In this case we have all three loss distributions with 95\%-VaRs of 0 and Maximum Losses of 5. Here the 95\%-Expected Loss will distinguish the loss distributions as  $ES(X_1) = \frac{5}{3}$, $ES(X_2) = \frac{5}{2}$, and $ES(X_3) = \frac{10}{3}$.\\

\begin{figure}[h!]
\centering%
\includegraphics{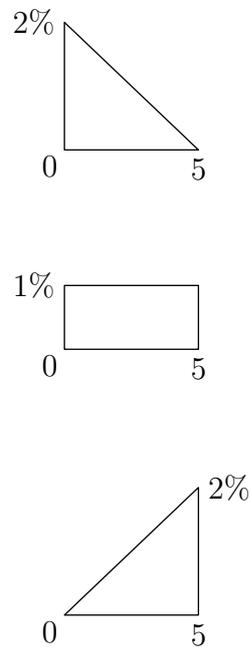}
\caption{Loss distributions with equal 95\%-VaRs and equal MLs}
\end{figure}

\noindent So now we will consider the final pair, the circumstance of using the 95\%-ES and the ML as our risk measure pair. \\

\begin{figure}[h!]
\centering%
\includegraphics{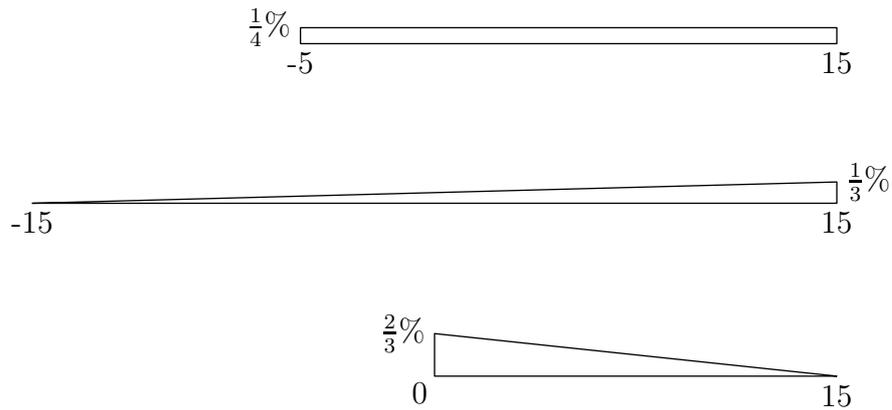}
\caption{Loss distributions with equal 95\%-ESs and equal MLs}
\end{figure}

\noindent Once again the three loss distributions shown in Figure 7 have contrasting risk situations, including a uniform probability, one of rising probabilities, and one of declining probabilities. Yet we have each of the three loss distributions with 95\%-ESs of 5 and with MLs of 15. The 95\%-VaR does differentiate among the distributions, as $VaR(X_1) = -5$, $VaR(X_2) = -15$, and $VaR(X_3) = 0$.

\subsection{Insufficiency of all three measures}

\noindent With the failure of two measures to definitively distinguish among different risk situations, we move to the next logical step. Here we look at the 95\%-VaR, the 95\%-ES, and the Maximum Loss.\\

\begin{figure}[h!]
\centering%
\includegraphics{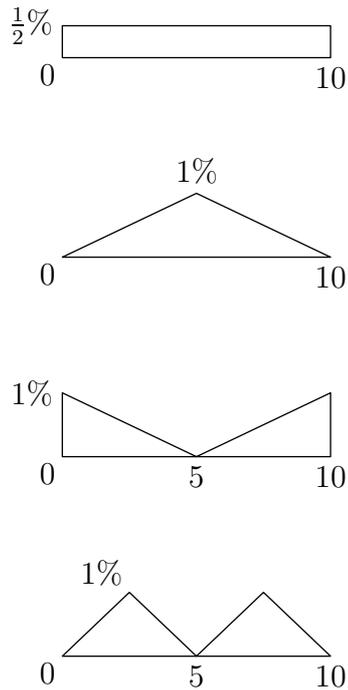}
\caption{Four loss distributions with equal 95\%-VaRs, 95\%-ESs and MLs}
\end{figure}

\noindent In Figure 8 all four of our examples have a 95\%-VaR of 0, a 95\%-ES of 5, and a ML of 10. The most compelling question here is whether these loss distributions are even possible in actual situations. We see uniform distributions with great regularity, so it should be clear that such a distribution is possible in a financial setting. The second loss distribution seems strange, but we argue that it is possible with properly chosen correlation. If we go back to the example of community banks making loans, we would expect a distribution that is very nearly normal. However, if this bank were to make loans in several communities, it is possible to see ramping probabilities in one neighborhood due to contagion. Then there may be other neighborhoods where most of the individuals have a large percentage of equity in their house. Such a situation could lead to the second loss distribution. And similar arguments can apply for the third distribution. Once that final neighborhood sees the blight of an entire region due to contagion of foreclosures, we could see that final neighborhood also experience increasing numbers of foreclosures. \\

\noindent The fourth loss distribution is nearly impossible for us to justify. However, as an academic exercise, it does give rise to a potentially infinite class of examples that share the same VaR, ES, and ML. In order for our examples to have equivalence of all three measures, we have to meet three criteria. We need the same ``starting points'' of the worst 5\% of returns in order to make the VaRs equal. We require identical ``ending points'' for the MLs to be equal. Since the uniform distribution has symmetry, it will have a 95\%-ES of 5. Thus, any distribution with an equivalent 95\%-ES (and the same starting and ending points) must also be symmetric in the greatest 5\% of its losses. So we could form a figure of $n$ triangles of height 1 and base $\frac{10}{n}$ on the interval $\lbrack 0,10 \rbrack$. \\

\noindent So our three measures do not distinguish among the potential scenarios that we might face. Obviously, something more is needed.

\subsection{Extending to different probability levels}

\noindent We have chosen to use the 95\% level for our risk measures, but there is nothing special about the 95\% level. Our examples would be just as valid if we substituted 90\% or 99\% or any other number in all of our measures and then scaled identically in all of the corresponding diagrams. However, this concept of considering different risk levels might also lead us to enough risk measures so that we can determine the universal risk profile from information about the different risk measures.\\

\noindent So now we will try using a vector of measures consisting of the 95\%-VaR, the 99\%-VaR, the 95\%-ES, 99\%-ES, and the ML. The different risk profiles demonstrated in Figure 8 are quite odd-looking, but they all have identical 95\%-VaRs, 99\%-VaRs, 95\%-ESs, 99\%-ESs, and MLs. \\

\begin{figure}[h!]
\centering%
\includegraphics{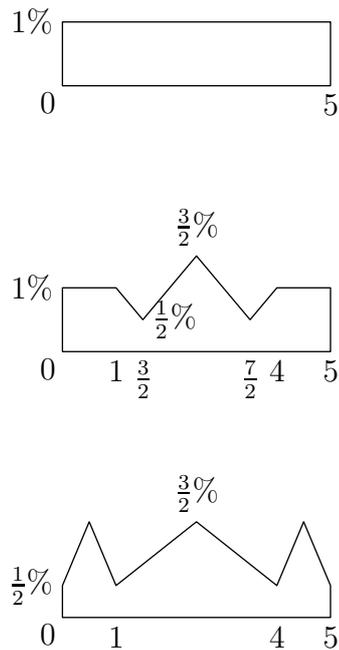}
\caption{Using five risk measures}
\end{figure}

\noindent The three distributions all have the necessary geometric properties for equality of all five measures. If we wanted to form a larger family of examples with the same values for all five measures, we would need our new distributions to have the same ``ending points'' in order to match up the MLs. We would need them to have the same ``starting points'' for both the greatest 5\% of losses and the greatest 1\% of losses in order to have equal 95\%-VaRs and equal 99\%-VaRs, respectively. And we would need symmetry in the greatest 5\% of losses and in the greatest 1\% of losses in order to have 95\%-ESs and 99\%-ESs, respectively, equal to the 95\%-ES and the 99\%-ES of the uniform distribution. Any figure meeting all these criteria would also be indistinguishable from the first five when viewed purely by looking at the values in our vector of five measures. Clearly several different profiles can share all five measures. \\

\noindent In spite of our effort in the academic exercise of drawing such figures, we do not claim that such loss distributions will occur naturally, with the obvious exception of the uniform distribution. Further, the loss distributions $X_2$ and $X_3$ do not differ by a great deal. Therefore, we are prepared to say that five risk measures should distinguish among differing loss distributions in almost all natural circumstances.

\section{Conclusion and Future Directions}

\noindent We have shown ambiguity among situations of rising and falling probability in loss distributions when one, two, and three risk measures are reported. Thus, we suggest that banks be required to submit five risk measures to regulatory bodies, specifically a 95\%-VaR, 99\%-VaR, 95\%-ES, and 99\%-ES, and a Maximum Loss, where this is possible. We have demonstrated that the reporting of five risk measures can theoretically lead to an infinite family of possibilities for the actual loss distribution, but we have also conceded that such loss distributions are unlikely to occur. Thus, we feel that banking regulations would be much safer if such regulations require the reporting of five different risk measures.\\

\noindent This is an interesting theoretical result that must be pursued on actual loss data. A paper that looks at the ability to differentiate between different types of common loss distributions is being written with my student, Cole Arendt. I am presently considering the question of how many risk measures are needed if we vary the timeframes on the historical data along with the level and the specific measures. Finally, another paper that looks at the predictive ability of historical risk measures is also in the works.\\

\end{document}